\documentclass[12pt]{article}
\usepackage{amsmath,amsfonts}
\def\beq{\begin{equation}}
\def\eeq{\end{equation}}
\def\bea{\begin{eqnarray}}
\def\eea{\end{eqnarray}}
\def\nn{\nonumber}

\makeatletter
 
  \def\@cite#1#2{${\mbox{#1\if@tempswa , #2\fi}}$}
\makeatother

\begin{document}
{\begin{center}
{\LARGE\sf An extension of the Bernoulli polynomials inspired by the Tsallis statistics}\\
\bigskip \bigskip\bigskip M. Balamurugan$^{\dagger}$, R. Chakrabarti$^{\ddagger}$\footnote{ranabir@imsc.res.in} and R. Jagannathan$^{\ddagger}$\footnote{jagan@cmi.ac.in}\\

\begin{small}
\bigskip
\textit{
$^{\dagger}$ Department of Theoretical Physics, 
University of Madras, \\
Maraimalai Campus, Guindy, 
Chennai, TN 600025, India \\}
 \textit{$^{\ddagger}$Chennai Mathematical Institute, H1 SIPCOT IT Park, \\ 
 Siruseri, Kelambakkam, TN 603103, India}\\
\end{small}
\end{center}
\vfill
\begin{abstract}
\noindent In [\cite{C1956}, \cite{C1979}] Carlitz introduced the degenerate Bernoulli numbers and polynomials by replacing the exponential factors in the corresponding classical  generating functions with their deformed analogs: $\exp(t) \rightarrow (1+\lambda t)^{1/\lambda}$, and 
$\exp(tx) \rightarrow (1+\lambda t)^{x/\lambda}$. The deformed exponentials reduce to their ordinary counterparts in the $\lambda \rightarrow 0$ limit. In the present work we study the extension of the Bernoulli polynomials obtained via an alternate deformation 
$\exp(tx) \rightarrow (1+\lambda tx)^{1/\lambda}$ that is inspired by the concepts of $q$-exponential function and $q$-logarithm used in the nonextensive Tsallis statistics.
\end{abstract}  

\newpage
\setcounter{page}{1}
\noindent 
\section{Introduction} 
\label{intro}

\medskip 

\noindent 
The well-known generating functions for the standard Bernoulli numbers $\{B_n\}_{n \geq0}$ and  Bernoulli polynomials 
$\{B_n(x)\}_{n \geq0}$ are, respectively, given by 
\beq
\frac{t}{\exp (t) - 1} = \sum_{n=0}^\infty B_n\frac{t^n}{n!}, \qquad 
\frac{t \exp (t x)}{\exp (t) - 1} = \sum_{n=0}^\infty B_n(x)\frac{t^n}{n!}, \qquad 
B_n(0) = B_n.
\label{bn-bp} 
\eeq 
Modifying the above generating functions via the replacement 
$\exp (t) \rightarrow (1+\lambda t)^{1/\lambda}$ for $\lambda \neq 0$,  Carlitz  introduced [\cite{C1956}, \cite{C1979}] the degenerate Bernoulli numbers $\{\beta_n(\lambda)\}_{n \geq0}$
\beq 
\frac{t}{(1+\lambda t)^{1/\lambda} - 1} = \sum_{n=0}^\infty \beta_n(\lambda)\frac{t^n}{n!},  
\label{betalambda} 
\eeq 
and the corresponding Bernoulli polynomials $\{\beta_n(\lambda ,x)\}_{n \geq0}$ as follows:
\beq 
\frac{t (1+\lambda t)^{x/\lambda}}{(1+\lambda t)^{1/\lambda} - 1} 
     = \sum_{n=0}^\infty \beta_n(\lambda ,x)\frac{t^n}{n!}, \qquad 
\beta_n(\lambda ,0) = \beta_n(\lambda). 
\label{betalambdax} 
\eeq 
Obviously, in the  $\lambda$ $\rightarrow$ $0$ limit the degenerate Bernoulli numbers and polynomials (\ref{betalambda}, \ref{betalambdax}) reduce, respectively, to  
their classical analogs.  

\par

On the other hand, the nonextensive statistical mechanics, pioneered by Tsallis [\cite{T2009}], is based  on a deformed exponential function, called the Tsallis $q$-exponential function, defined as 
\beq
\exp_q(\mathcal{X}) = (1+(1-q) \mathcal{X})^{1/(1-q)},  \qquad 
\lim_{q \rightarrow 1}\; \exp_q(\mathcal{X}) = \exp (\mathcal{X}).
\label{qexp}
\eeq  
The deformed exponential (\ref{qexp}) may be inverted via the corresponding  $q$-logarithm:   
\beq 
\log_q(\mathcal{X}) = \frac{\mathcal{X}^{1-q}-1}{1-q}, \qquad 
\log_q(\exp_q(\mathcal{X})) = \mathcal{X}.
\label{lnq} 
\eeq 
Identifying the parameters $(1-q) \rightarrow \lambda$ it readily follows that the Tsallis $q$-exponential function (\ref{qexp}) is precisely same as the modification introduced  by Carlitz to construct the generating relations (\ref{betalambda}, \ref{betalambdax}). For our purpose, we slightly alter the notations given in (\ref{qexp}, \ref{lnq}):
\beq
\exp_\lambda(\mathcal{X}) = (1+\lambda \mathcal{X})^{1/\lambda},\quad 
\log_\lambda(\mathcal{X}) = \frac{1}{\lambda}(\mathcal{X}^\lambda - 1),
\label{lambdaexplog}
\eeq
where the inversion relation reads: $\log_\lambda(\exp_\lambda(\mathcal{X})) = \mathcal{X}$. The inverse relationship between the $\lambda$-logarithm and the $\lambda$-exponential functions are, however, not utilized in the choice (\ref{betalambdax}): 
\beq 
\log_\lambda((\exp_\lambda(t))^x) =  \frac{1}{\lambda}((1+\lambda t)^x - 1) 
     = \sum_{n = 1}^{\infty} \lambda^{n-1}\, (x)_n\, \frac{t^n}{n!}, 
\label{log-E-noninv}
\eeq
where the falling factorial reads $(x)_n=x(x-1)(x-2)\cdots(x-n+1)$. On the other hand, the alternate deformation
\beq
\exp(t x) \rightarrow \exp_\lambda(t x)
\label{alt-de}
\eeq
employs the inverse functional property: $\log_{\lambda}(\exp_{\lambda} (t x)) = t x$. The above discussion  illustrates that if we replace $\exp(t x)$ in the generating function (\ref{bn-bp}) by its $\lambda$-analog then the two substitutes, namely, $(\exp_{\lambda}(t))^x$ and 
$\exp_{\lambda} (t x)$, despite both of them  agreeing in the limit $\lambda \rightarrow 0$, lead to drastically different properties. Contrasting the choice  followed in  [\cite{C1979}], we, in 
this work, explore some of the consequences of introducing the other deformation (\ref{alt-de}) 
 in the generating relation (\ref{bn-bp}) for the Bernoulli polynomials.

\section{$\widetilde{\beta}$-Bernoulli polynomials} 
\label{beta-bernoulli}
\subsection{Generating function}
\setcounter{equation}{0} 
We introduce the deformed Bernoulli polynomials $\{\widetilde{\beta}_n(\lambda|x)\}_{n \geq0}$ by altering the generating function  (\ref{betalambdax}) as follows:
\beq 
\frac{t \,\exp_\lambda(t x)}{\exp_\lambda(t) - 1} 
     = \sum_{n=0}^\infty \widetilde{\beta}_n(\lambda| x)\frac{t^n}{n!},\quad 
\widetilde{\beta}_n(\lambda| 0) = \beta_n(\lambda), \quad
\lim_{\lambda \rightarrow 0} \widetilde{\beta}_n(\lambda| x) = B_n(x).
\label{tildebetax} 
\eeq 
Substituting the binomial expansion  
\beq 
\left(\exp_\lambda(\mathcal{X})\right)^{\pm 1}
                   = \sum_{n=0}^\infty (\pm 1)^n \varepsilon^{(\mp)}_\lambda(n) \frac{\mathcal{X}^n}{n!}, \quad
\varepsilon_\lambda^{(\pm)}(n) = \left\{ \begin{array}{ll} 
                                            1, & \mathrm{for}\ n = 0, \\
                                            \prod_{\jmath =0}^{n-1} (1\pm \jmath\lambda), & \mathrm{for}\ n\geq 1 
                                            \end{array} \right.                   
\label{elambdaseries} 
\eeq 
in the generating function  (\ref{tildebetax}), we obtain 
\beq 
 \sum_{n=0}^\infty \varepsilon^{(-)}_\lambda(n)\frac{t^n x^n}{n!} = \left(\sum_{k=0}^\infty 
 \varepsilon^{(-)}_\lambda(k+1)\frac{t^k}{(k+1)!}\right)
   \left(\sum_{\ell=0}^\infty \widetilde{\beta}_\ell(\lambda| x)\frac{t^\ell}{\ell!}\right).
\label{xn-lambda-expn}     
\eeq 
Equating terms $O(t^{n})$ on both sides in (\ref{xn-lambda-expn})  we now produce the expansion of $x^n$ via the $\widetilde{\beta}_n(\lambda|x)$ polynomials: 
\beq 
\varepsilon^{(-)}_\lambda(n)\, x^n = \frac{1}{n+1} \sum_{k=0}^n \left( \begin{array}{c} n+1 \\ k \end{array} \right)\, 
\varepsilon^{(-)}_\lambda(n+1-k)\, \widetilde{\beta}_k(\lambda| x), \qquad n \geq 0. 
\label{tildebetaxrecrel}   
\eeq
The above equation directly furnishes the recurrence formula:
\beq
\widetilde{\beta}_n(\lambda| x) = \varepsilon^{(-)}_\lambda(n)\, x^n -  \frac{1}{n+1} \sum_{k=0}^{n- 1} \left( \begin{array}{c} n+1 \\ k \end{array} \right)\, \varepsilon^{(-)}_\lambda(n+1-k)\, \widetilde{\beta}_k(\lambda| x).
\label{ber-x-recur}
\eeq
The generating function (\ref{tildebetax}) also provides an explicit construction of the  $\widetilde{\beta}_n(\lambda|x)$ polynomials 
in terms of the $\lambda$-Bernoulli numbers given in (\ref{betalambda}):
\beq
 \widetilde{\beta}_{n}(\lambda| x) =\sum_{k= 0}^{n} \binom{n}{k} \varepsilon_{\lambda}^{(-)}(k)\, 
 \beta_{n - k}(\lambda)\, x^{k}, \qquad n \geq 0. 
 \label{beta-x}
\eeq
It is interesting to note that, \textit{a \`{l}a} Carlitz [\cite{C1961}], the above expansion may be symbolically expressed as 
\beq
 \widetilde{\beta}_{n}(\lambda| x) = (\beta(\lambda) + \varepsilon_{\lambda}^{(-)}\, x)^{n},
 \label{SymbolicExpn}
 \eeq
where it is understood that after expansion of the right member we replace $\beta(\lambda)^{k}$ with $\beta_{k}(\lambda)$, and $\varepsilon_{\lambda}^{(-)\,k}$ with $\varepsilon_{\lambda}^{(-)}(k)$.  Following the above expansion we now list the first few  $\widetilde{\beta}_n(\lambda|x)$ polynomials as follows:
\bea
\widetilde{\beta}_0(\lambda| x) & = & 1, \; \widetilde{\beta}_1(\lambda| x) =  x + \frac{1}{2} (\lambda - 1), \;
\widetilde{\beta}_2(\lambda| x) = \varepsilon_{\lambda}^{(-)}(2)\,  x^2 + (\lambda - 1) x - \frac{1}{6} (\lambda^2 - 1), \nn \\ 
\widetilde{\beta}_3(\lambda| x) &=& \varepsilon_{\lambda}^{(-)}(3)\, x^3 -\frac{3}{2} (1-\lambda)^2 x^2   
+\frac{1}{2} \left(1-\lambda^{2}\right) x - \frac{\lambda}{4} \left(1-\lambda^2 \right), \nn \\
\widetilde{\beta}_4(\lambda| x) &=&  \varepsilon_{\lambda}^{(-)}(4)\, x^4 -2(1-\lambda)^2(1-2\lambda)x^3 
+ (1-\lambda)(1-\lambda^2)x^2-\lambda (1-\lambda^2) x\nn\\ 
&&- \frac{1}{30} (1-20\lambda^2+19 \lambda^4). 
\label{ber01234}
\eea 

\par

Proceeding further we now develop a translation of the variable of the $\widetilde{\beta}_n(\lambda|x)$ polynomials. Towards this we note that the generating function (\ref{tildebetax}) may be written in a factorized form: 
\beq 
\frac{t \,\exp_\lambda(t (x+y) )}{\exp_\lambda(t) - 1} =  \frac{t \,\exp_\lambda(t x)}{\exp_\lambda(t) - 1} \frac{\exp_\lambda(t (x+y) )}{\exp_\lambda(t x)}.
\label{gen-fac}
\eeq
Utilizing the binomial expansion (\ref{elambdaseries}) we obtain the desired expression for a translational shift in the variable:
\beq
 \widetilde{\beta}_{n}(\lambda| x+y) = \sum_{k, \ell = 0}^{n} (-1)^{k} \,\binom{n}{k, \ell, n -k - \ell} \varepsilon_{\lambda}^{(+)}(k)\, 
 \varepsilon_{\lambda}^{(-)}(\ell)\, \widetilde{\beta}_{n - k - \ell}(\lambda| x)\, x^{k}\, (x +y)^{\ell},
 \label{x+y}
\eeq
where the multinomial coefficient is given by $\binom{n}{k, \ell, n -k - \ell}= \frac{n!}{k! \ell! (n -k - \ell)!}$.  We observe that a contraction in the variables $x \rightarrow 0, y \rightarrow x$ in (\ref{x+y}) readily yields the expansion (\ref{beta-x}) of the $\widetilde{\beta}_n(\lambda|x)$ polynomials directly obtained earlier. Moreover, the classical limit $\lambda \rightarrow 0$ of the translation property (\ref{x+y}) may be realized by noting the reduction of the coefficients 
$\varepsilon^{(\pm)}_\lambda(n)|_{\lambda \rightarrow 0} \rightarrow 1$. 
Implementation of the above limits in (\ref{x+y}) now leads to the well-known shift property of the Bernoulli polynomials:
\beq
 B_n(x+y) = \sum_{\ell=0}^n \binom{n}{\ell} B_{n- \ell}(x)\, y^{\ell}.
\label{B-shift}
\eeq

\subsection{$\lambda$-Appell sequence}
\label{Appell}
The  sequence $\{P_n(x)\}_{n \geq 0}$ of Appell polynomials [\cite{A1882}] satisfy the property
\beq
f(t)\, \exp(x g(t)) = \sum_{n=0}^\infty P_n(x)\frac{t^n}{\phi(0)\ldots\phi(n)}, \qquad f(0) \ne 0, g(0)= 0, g'(0) \ne 0,
\label{appell}
\eeq
where $\phi: \mathbb{N} \rightarrow \mathbb{C}/\{0\}$ is an arbitrary function. The generating function introduced in (\ref{tildebetax}) suggests a slight generalization of the above relation. In particular, we replace the exponential function in the l.h.s. of (\ref{tildebetax}) possessing 
a factorized form of the argument with a more general construct: 
\beq
\exp(x g(t)) \rightarrow \exp(\mathfrak {g} (x, t)), 
\mathrm{where} \;\;\mathfrak {g} (x, 0) = 0, 
\tfrac{\mathrm{d} \mathfrak {g} (x, t)}{\mathrm {d} t}|_{t = 0} \ne 0.
\label{gxt}
\eeq
As the $\lambda$-deformed exponential admits an infinite product representation
\beq
\exp_\lambda(t x) = \exp \left\lgroup \sum_{\ell = 0}^{\infty} (-1)^{\ell} \lambda^{\ell}\,\tfrac{(t x)^{\ell + 1}}{\ell + 1}
\right\rgroup,
\label{infinit-prod}
\eeq
the generating function structure (\ref{tildebetax}) provides an example of the generalized Appell sequence discussed above. 

\par

The Bernoullli polynomials form a well-known Appell sequence. As a consequence it obeys  the recursive property: $B'_n(x) = nB_{n-1}(x)$ that we attempt to generalize now. Towards this end we first  note the eigenfunction structure
\beq 
 \mathcal{D}_{\lambda}(x) \exp_\lambda(tx) = t \,\exp_\lambda(tx), \; \mathcal{D}_{\lambda}(x) \equiv (1+\lambda tx)\,\frac{{\mathrm d}}{{\mathrm d}x}
\label{d-x-exp}
\eeq 
and subsequently operate  both sides of the defining relation for the $\widetilde{\beta}$-Bernoulli polynomials (\ref{tildebetax}) from the left by the composite derivative $\mathcal{D}_{\lambda}(x)$:
\beq 
(1+\lambda tx)\sum_{n=0}^\infty \widetilde{\beta}'_n(\lambda| x)\frac{t^n}{n!} 
    = \sum_{n=0}^\infty \widetilde{\beta}_n(\lambda| x)\frac{t^{n+1}}{n!}. 
\label{app-exp}    
\eeq  
Employing $\widetilde{\beta}'_0(\lambda|x) = 0$, the above equation may be recast  as 
\beq 
\sum_{n=1}^\infty \widetilde{\beta}'_n(\lambda| x)\frac{t^n}{n!} 
        = \sum_{n=1}^\infty n\widetilde{\beta}_{n-1}(\lambda| x)\frac{t^n}{n!} 
                 - \lambda x \sum_{n=1}^\infty n\widetilde{\beta}'_{n-1}(\lambda| x)\frac{t^n}{n!}. 
\label{app-sum}                  
\eeq   
This leads to the desired result: 
\beq 
\widetilde{\beta}'_n(\lambda| x) = n(\widetilde{\beta}_{n-1}(\lambda| x) - \lambda x \widetilde{\beta}'_{n-1}(\lambda| x)) 
\label{dtildebetandx} 
\eeq
that may be regarded as the defining characteristic of the $\lambda$-Appell sequence. In the classical  $\lambda$ $\rightarrow$ $0$ limit it reduces to the standard structure. The recurrence relation (\ref{dtildebetandx}) admits a series solution for the derivatives of the polynomials $\widetilde{\beta}_{n}(\lambda| x)$:
\beq 
\widetilde{\beta}'_{0}(\lambda| x) = 0,\quad
\widetilde{\beta}'_{n+1}(\lambda| x) = (n+1)!\, \sum_{k = 0}^{n}\, \frac{(- \lambda x)^{n - k}}{k!}\,\widetilde{\beta}_{k}(\lambda| x).  
\label{Bder_B} 
\eeq

\par
 
Equation (\ref{dtildebetandx}) shows that the $\widetilde{\beta}$-Bernoulli polynomials form a $\lambda$-Appell sequence constructed in the following way: 
\beq
\widetilde{\beta}_n(\lambda| x) = \widetilde{\beta}_n(\lambda| 0) + n\int_0^x (\widetilde{\beta}_{n-1}(\lambda| u) - \lambda u
 \widetilde{\beta}'_{n-1}(\lambda| u))\,\mathrm {d}u, \qquad n = 1,2,\cdots.
\label{lambdaB}  
\eeq 
In general, a sequence of polynomials $\{P_n(\lambda| x)_{n \geq 0}\}$ may be said to form a 
$\lambda$-Appell sequence if 
\beq
P_n(\lambda| x) = P_n(\lambda| 0) + n\int_0^x (P_{n-1}(\lambda| s) - \lambda s P'_{n-1}(\lambda| s))ds, \qquad n = 1,2,\cdots. 
\label{lambdaP}
\eeq 
For example, letting $P_n(\lambda| 0) = 0$, we see that 
$\{ \varepsilon_\lambda (n) x^n, \ n = 0,1,2,\cdots \}$ form a $\lambda$-Appell sequence. 

\subsection{Determinant structure}
Recently a determinantal approach for the Bernoulli polynomials $B_{n}(x)$ has been proposed [\cite{C1999}, \cite{CDG2006}]. The construction employs an upper Hessenberg matrix [\cite{HJ1985}] with the entries  $h_{\jmath, \ell} = 0$ if $\jmath - \ell \ge 2$. The determinant 
of an upper Hessenberg matrix of order $n$  
\beq
H_n 
= 
\begin{vmatrix}
h_{1,1} & h_{1,2} & h_{1,3} & \cdots & \cdots & \cdots & h_{1,n} \\ 
h_{2,1} & h_{2,2} & h_{2,3} & \ddots & \ddots & \ddots &h_{2,n} \\
0 & h_{3,2} & h_{3,3} & \ddots & \ddots & \ddots &h_{3,n} \\
\vdots & 0 & h_{4,3} & h_{4,4} & \ddots & \ddots &h_{4,n} \\
\vdots & & \ddots & \ddots & \ddots & \ddots & \vdots \\
\vdots & & \ddots & \ddots & \ddots & \ddots & \vdots  \\
0 & \cdots & \cdots & \cdots  & 0 & h_{n,n-1} & h_{n,n},
\end{vmatrix}
\label{Hmatrix}
\eeq
 obeys the following recursive relation [\cite{CDG2006}]:
\beq
H_{0} \equiv 1,\qquad H_{n(\ge 1)} = \sum_{\ell = 0}^{n-1} (-1)^{n-1-\ell} \,\mathcal{Q}_{\ell}^{n-2}\,
h_{\ell+1, n}\,H_{\ell},
\label{H-recurrence}
\eeq
where the factorized coefficients read: $\mathcal{Q}_{\ell (\le k)}^{k} = \prod_{\jmath = \ell}^{k} h_{\jmath +2,\, \jmath+1},\,
\mathcal{Q}_{\ell (> k)}^{k} \equiv 1$. Employing the above recurrence relation we now provide the representation of the polynomials 
$\widetilde{\beta}_n (\lambda|x)$ via a $(n+1) \times (n+1)$ determinant. Towards this end we define
\beq
\widetilde{\beta}_{0} (\lambda|x)=1,\qquad\widetilde{\beta}_{n (\ge 1)} (\lambda|x)
= \frac{(-1)^n}{(n-1)!} \mathsf{D}_n(\lambda|x),
\label{Ber-det}
\eeq
where the determinantal function $\mathsf{D}_n(\lambda|x)$ reads
\bea
\begin{vmatrix}
1 & \! \! \varepsilon_{\lambda}^{(-)}(1) x & \varepsilon_{\lambda}^{(-)}(2) x^2 & \varepsilon_{\lambda}^{(-)}(3) x^3 & \! \! \! \! \cdots & 
\! \! \! \! \varepsilon_{\lambda}^{(-)}(n-1)x^{n-1} &\! \!  \varepsilon_{\lambda}^{(-)}(n) x^n \\ 
\varepsilon_{\lambda}^{(-)}(1) & \varepsilon_{\lambda}^{(-)}(2) \, \frac{1}{2} & \varepsilon_{\lambda}^{(-)}(3) \,  \frac{1}{3} & 
\varepsilon_{\lambda}^{(-)}(4) \, \frac{1}{4}& \! \! \! \! \cdots & \varepsilon_{\lambda}^{(-)}(n) \,  \frac{1}{n} &\! \!  \varepsilon_{\lambda}^{(-)}(n+1) \, \frac{1}{n+1} \\
0 & \! \! \varepsilon_{\lambda}^{(-)}(1) & \varepsilon_{\lambda}^{(-)}(2) & \varepsilon_{\lambda}^{(-)}(3) & 
\! \! \! \! \cdots & \! \! \! \! \varepsilon_{\lambda}^{(-)}(n-1) &\! \!  \varepsilon_{\lambda}^{(-)}(n) \\
0 & \! \!  0 & \varepsilon_{\lambda}^{(-)}(1) \, 2 & \varepsilon_{\lambda}^{(-)}(2) \, 3 & \! \! \! \! \cdots &
 \! \! \! \! \varepsilon_{\lambda}^{(-)}(n-2) \, (n-1) &\! \!  \varepsilon_{\lambda}^{(-)}(n-1) \, n \\
0 & \! \!  0 & 0 &  \varepsilon_{\lambda}^{(-)}(1) \binom{3}{2} & \! \! \! \! \cdots & \! \! \! \! \varepsilon_{\lambda}^{(-)}(n-3) \, \binom{n-1}{2}
 & \! \! \varepsilon_{\lambda}^{(-)}(n-2) \, \binom{n}{2} \\
\vdots &\! \!  \vdots & \vdots & \vdots & \! \! \! \! \ddots & \! \! \! \! \vdots & \vdots  \\
0 & 0 & 0 & 0  & \! \! \! \! \cdots & \! \! \! \! \varepsilon_{\lambda}^{(-)}(1) \, \binom{n-1}{n-2} & \! \!  \varepsilon_{\lambda}^{(-)}(2) \, \binom{n}{n-2}
\end{vmatrix}.
\label{D-Ber}
\eea
The recurrence relation (\ref{H-recurrence}) utilized for the above Hessenberg determinant $\mathsf{D}_n(\lambda|x)$ reproduces the defining property (\ref{ber-x-recur}) of the polynomials $\widetilde{\beta}_{n} (\lambda|x)$. In the classical $\lambda \rightarrow 0$ limit the above construction agrees to structure given in [\cite{CDG2006}].

\subsection{A two variable generalization}
The two variable Kamp\`{e} de F\`{e}ri\`{e}t generalization of the Hermite polynomials  has been introduced [\cite{B1934}, \cite{DLC1999}] via an extension of the generating function. Employing this method a new class  of the Bernoulli polynomials has been obtained [\cite{DCL2002}]. This technic has been recently used [\cite{KRA2016}] to establish certain mixed special polynomial families associated with Appell sequences. In the context of Tsallis type of deformation of the Bernoulli polynomials introduced here we now attempt the corresponding two variable extension via the following  factorized generating function:
\beq 
\frac{t}{\exp_\lambda(t) - 1} \,\exp_\lambda(t x) \,\exp_\lambda(t^{r} y) = \sum_{n=0}^\infty 
\widetilde{\beta}_{n}^{(r)}(\lambda| x, y)\,\frac{t^n}{n!},
\quad \widetilde{\beta}_{n}^{(r)}(\lambda| x, 0) = \widetilde{\beta}_{n}(\lambda| x). 
\label{tildebetaxy} 
\eeq 
The recurrence relation for the $\widetilde{\beta}_{n}^{(r)}(\lambda| x, y)$ polynomials follows by equation the coefficients of the identical powers of $t^{n}$ on both sides of (\ref{tildebetaxy}):
\bea
\widetilde{\beta}_{n}^{(r)}(\lambda| x, y) &=& \sum_{\ell =0}^{\lfloor\frac{n}{r}\rfloor}\frac{n!}{(n-r \ell)!\,\ell!}
\varepsilon^{(-)}_\lambda(\ell)\,\varepsilon^{(-)}_\lambda(n- r \ell)\, x^{n-r \ell}\,y^{\ell}\nn\\
&& -  \frac{1}{n+1} \sum_{k=0}^{n- 1} \binom{n+1}{\ell}\, \varepsilon^{(-)}_\lambda(n+1-\ell)\, \widetilde{\beta}_{\ell}^{(r)}
 (\lambda| x, y).
\label{ber-xy-recur}
\eea
We note that a null choice $y = 0$ in (\ref{ber-xy-recur}) readily reproduces the recurrence relation (\ref{ber-x-recur}) observed for the single variable case. The recurrence relation (\ref{ber-x-recur}) admits explicit solution of the polynomials 
$\widetilde{\beta}_{n}^{(r)}(\lambda| x, y)$ as a double sum:
\bea
 \widetilde{\beta}_{n(\ge 0)}^{(r)}(\lambda| x, y) &=& \sum_{\jmath= 0}^{n} \sum_{\ell =0}^{\lfloor\frac{n}{r}\rfloor} 
 \frac{n!}{\jmath ! \,{\ell}!\, 
 (n- \jmath -r \ell)!} \;\varepsilon_{\lambda}^{(-)}(\jmath)\, \varepsilon_{\lambda}^{(-)}(\ell)\,
 \beta_{n - \jmath -r \ell}(\lambda)\, x^{\jmath}\,y^{\ell}\nn\\
 &=&\sum_{\ell =0}^{\lfloor\frac{n}{r}\rfloor} \frac{n!}{\ell ! \,
 (n -r \ell)!}\, \varepsilon_{\lambda}^{(-)}(\ell)\,
 \widetilde{\beta}_{n -r \ell}(\lambda| x)\, y^{\ell},
 \label{beta-xy}
\eea
where the second equality provides the connection formula between the two variable and the single variable polynomials. The eigenfunction structure of the deformed exponential
\beq 
 \mathcal{D}_{\lambda}(y) \exp_\lambda(t^{r} y) = t^{r} \,\exp_\lambda(t^{r} y), \; \mathcal{D}_{\lambda}(y) 
 \equiv (1+\lambda\, t^{r} y)\,\frac{{\partial}}{{\partial} y}
\label{d-y-exp}
\eeq 
allows us to observe the differential recurrence properties of the polynomials $\widetilde{\beta}_{n}^{(r)}(\lambda| x, y)$:
\bea 
\frac{\partial \widetilde{\beta}_{n}^{(r)}(\lambda| x, y)}{\partial x} &=& n \left(\widetilde{\beta}_{n-1}^{(r)}(\lambda| x, y) - 
\lambda\, x \,\frac{\partial \widetilde{\beta}_{n-1}^{(r)}(\lambda| x, y)}{\partial x}\right), 
\label{x-boundary} \\
\frac{\partial \widetilde{\beta}_{n}^{(r)}(\lambda| x, y)}{\partial y} &=& \frac{n!}{(n-r)!}\, \left(\widetilde{\beta}_{n-r}^{(r)}
(\lambda| x, y) - \lambda\, y \,\frac{\partial \widetilde{\beta}_{n-r}^{(r)}(\lambda| x, y)}{\partial y}\right). 
\label{y-boundary} 
\eea
The differential structure (\ref{x-boundary}, \ref{y-boundary}) provides the defining relations of the extension of $\lambda$-Appell sequences for the two variable polynomials.

\section{Conclusion}
Altering the generating function of the Bernoulli polynomials via the Tsallis  exponential function we have introduced a new deformation  $\widetilde{\beta}_{n} (\lambda|x)$ of the said polynomials. These polynomials may be regarded as elements of the deformed $\lambda$-Appell sequence. Explicit expansion of the $\widetilde{\beta}_{n} (\lambda|x)$ polynomials and the corresponding translation formula are derived. A representation of these polynomials via a Hessenberg type of determinantal structure has been provided. By augmenting the factorized generating function  a two variable extension of these polynomials has been realized. Similar $\lambda$-deformations of, in particular, the Euler and Hermite polynomials may be of interest. A potential application of the polynomials discussed here lies in the area of nonextensive statistical mechanics [\cite{T2009}] where the Tsallis exponential function  plays a key role in the description of the entropy of the physical system. The polynomials presented here may be useful in developing a perturbative procedure for evaluation of such nonextensive statistical quantities. In particular, the investigations on the $(q -1)$ expansion [\cite{H2009}] of various physical quantities valid asymptotically, may be facilitated by employing the $\widetilde{\beta}_{n}(\lambda| x)$ polynomials.

%%%
\end{document}